# Relativistic Energies and Scattering Phase Shifts for the Fermionic Particles Scattered by Hyperbolical potential with the Pseudo (Spin) Symmetry

Oyewumi[+1], K. J and Oluwadare[+2], O. J

[+1]Department of Physics, Federal University of Technology, Minna, Niger State, Nigeria.
[+2]Department of Physics, Federal University Oye-Ekiti, P. M. B. 373, Ekiti State, Nigeria

**Abstract**

In this paper, we studied the approximate scattering state solutions of the Dirac equation with the hyperbolical potential with pseudospin and spin symmetries. Using a suitable short range approximation within the formalism of functional analytical method, we obtained the spin-orbit quantum numbers dependent scattering phase shifts for the spin and pseudospin symmetries. The normalization constants, lower and upper radial spinor for the two symmetries and the relativistic energy spectra were presented. Our results reveal that both the symmetry constants ($C_{ps}$ and $C_s$) and the spin-orbit quantum number κ affect scattering phase shifts significantly.

## 1.0 Introduction

Scattering theory is very central to the study of several of fields such as atomic, nuclear, high energy, or condensed matter physics. It allows for descriptions and interpretations of many collisions process such as excitation and ionization by particle or radiation impact, etc. [1-7]. Complete information about the quantum systems can only be obtained by investigating scattering state solutions of relativistic and non-relativistic equations with quantum mechanical potential model.

As a result, several authors in quantum mechanics have strictly followed different approaches to study the scattering state solutions of the relativistic and non-relativistic wave equations for central and non- central potential models [6-37]. In their works, they have reported the calculations on phase shifts, transmission and reflection coefficients, resonances, normalized radial wave functions and properties of S-matrix for potential models of their interest. All these are sufficient enough to predict, correlate and describes the behaviour of particles.

[+1]kjoyewumi66@unilorin.edu.ng;
[+1]On Sabbatical Leave from: Theoretical Physics Section, Department of Physics, University of Ilorin, Ilorin, Nigeria.
[+2]Corresponding author: oluwatimilehin.oluwadare@fuoye.edu.ng

In view of the above works, we are motivated to investigate the scattering state solutions of Dirac equation with hyperbolical potential by applying a suitable approximation within the formalism of functional analytical method.

This paper is organized as follows: Section 2 contains the basic equation, In Section 3, we studied the approximate scattering state solutions for the Hyperbolical potential in detail. In Section 4, the numerical results were presented and discussed. Finally, the conclusions are given in Section 5.

**2.0    The Basic Equations**

By considering the Dirac wave equation and its corresponding spinors, the two-coupled first order differential equations for the upper and lower components of the spinor may be obtained as [37-41]:

$$\left[\frac{d}{dr} + \frac{\kappa}{r}\right] F_{n\kappa}(r) = [M + E_{n\kappa} - \Delta(r)] G_{n\kappa}(r), \tag{1}$$

$$\left[\frac{d}{dr} - \frac{\kappa}{r}\right] G_{n\kappa}(r) = [M - E_{n\kappa} + \Sigma(r)] F_{n\kappa}(r), \tag{2}$$

where $\Delta(r) = V(r) - S(r)$ and $\Sigma(r) = V(r) + S(r)$. Solving for $G_{n\kappa}(r)$ in Eq. (1) and $F_{n\kappa}(r)$ in Eq. (2), we obtain the Schrödinger-like equations satisfying for upper radial spinor $F_{n\kappa}(r)$ and lower radial spinor $G_{n\kappa}(r)$, respectively as:

$$\left\{\frac{d^2}{dr^2} - \frac{\kappa(\kappa+1)}{r^2} + \left[-\left(M + E_{n\kappa} - \Delta(r)\right)\left(M - E_{n\kappa} + \Sigma(r)\right) + \frac{\frac{d\Delta(r)}{dr}\left(\frac{d}{dr} + \frac{\kappa}{r}\right)}{M + E_{n\kappa} - \Delta(r)}\right]\right\} F_{n\kappa}(r) = 0, \tag{3}$$

$$\left\{\frac{d^2}{dr^2} - \frac{\kappa(\kappa-1)}{r^2} + \left[-\left(M + E_{n\kappa} - \Delta(r)\right)\left(M - E_{n\kappa} + \Sigma(r)\right) + \frac{\frac{d\Sigma(r)}{dr}\left(\frac{d}{dr} - \frac{\kappa}{r}\right)}{M - E_{n\kappa} + \Sigma(r)}\right]\right\} G_{n\kappa}(r) = 0, \tag{4}$$

where $\kappa(\kappa - 1) = \tilde{l}(\tilde{l} + 1)$ and $\kappa(\kappa + 1) = l(l + 1)$.

**2.1    Pseudospin symmetry limit for the hyperbolical potential**

By following the pseudospin symmetry conditions and considering the hyperbolical potential $\Delta(r)$ satisfying relativistic model [42-46]:

$$\Delta(r) = V(r) = D[1 - \sigma_0 \coth(\alpha r)]^2, \tag{5}$$

where D, $\alpha$ and $\sigma_0$ are the three positive potential parameters that significantly affect the relativistic energy spectra and the relativistic scattering phase shifts. Schiöberg [43] reported that this potential is closely related to the Morse, the Kratzer, the Coulomb, the harmonic oscillators and other potential functions in a particular limits. The properties and applications of this potential have been given by Lu et al. (2005) and Schiöberg (1986).

Under the pseudospin symmetry condition Eq. (4) yields

$$\left\{\frac{d^2}{dr^2} - \frac{\kappa(\kappa-1)}{r^2} - \gamma + \beta D\left[1 - \frac{\sigma_0(1+e^{-2\alpha r})}{1-e^{-2\alpha r}}\right]^2\right\} G_{ps,n\kappa}(r) = 0, \qquad (6)$$

where $\gamma = (M + E_{n\kappa})(M - E_{n\kappa} + C_{ps})$ and $\beta = M - E_{n\kappa} + C_{ps}$ are the pseudospin symmetry energy parameters.

### 2.2 Spin symmetry limit for the hyperbolical potential

In a similar way, we consider the spin symmetry conditions and take $\Sigma(r)$ as hyperbolical potential [42-46]. i.e.

$$\Sigma(r) = V(r) = D[1 - \sigma_0 \coth(\alpha r)]^2, \qquad (7)$$

and using the spin symmetry conditions, Eq. (3) becomes

$$\left\{\frac{d^2}{dr^2} - \frac{\kappa(\kappa+1)}{r^2} - \tilde{\gamma} - \tilde{\beta} D\left[1 - \frac{\sigma_0(1+e^{-2\alpha r})}{1-e^{-2\alpha r}}\right]^2\right\} F_{s,n\kappa}(r) = 0, \qquad (8)$$

where $\tilde{\gamma} = (M - E_{s,n\kappa})(M + E_{s,n\kappa} - C_s)$ and $\tilde{\beta} = (M + E_{s,n\kappa} - C_s)$ denote spin symmetry energy parameters.

### 2.3 Pekeris-type approximation

In all the limits we apply the following short range approximation [47-52]:

$$\frac{1}{r^2} \approx 4\alpha^2 \left[c_0 + \frac{e^{-2\alpha r}}{1-e^{-2\alpha r}} + \left(\frac{e^{-2\alpha r}}{1-e^{-2\alpha r}}\right)^2\right], \qquad (9)$$

where $c_0 = 0.0823058167837972$ is a proper shift and when $c_0 = 0$, this approximation is identical to the one suggested by Greene & Aldrich [47]. The choice of this approximation is based on the fact that it is good approximation to $\frac{1}{r^2}$ term and it is very accurate for small potential parameter $\alpha$ and $\sigma_0$. Falaye [45] reported that the difference between the exact results and the one obtained by this approximation is very small and it can be ignored. Our interest here is to apply this approximation and see whether the symmetry constants ($C_{ps}$ and $C_s$) and the spin-orbit quantum number $\kappa$ will have any influence on the scattering phase shifts.

### 3. Relativistic scattering state solutions

### 3.1. Pseudospin Symmetry limit for hyperbolical potential

Defining a variable $z = 1 - e^{-2\alpha r}$ and applying approximation in Eq. (9), then, Eq. (6) reduces to

$$\left\{\frac{d^2}{dz^2} - \frac{1}{(1-z)}\frac{d}{dz} + \left[\frac{Pz^2 + Qz + R}{z^2(1-z)^2}\right]\right\} G_{ps,n\kappa}(r) = 0, \qquad (10)$$

with the following useful definitions:

$$P = \frac{k_1^2}{4\alpha^2} + \frac{\beta D \sigma_o}{\alpha^2}, \quad Q = \frac{4\alpha^2 \kappa(\kappa-1)c_0 - 4\beta D \sigma_0(1+\sigma_0)}{4\alpha^2}, \quad R = \frac{-4\alpha^2 \kappa(\kappa-1)c_0 + 4\beta D \sigma_0^2}{4\alpha^2}, \tag{11}$$

where $k_1 = \sqrt{\beta D(1-\sigma_0)^2 - \gamma - 4\alpha^2 \kappa(\kappa-1)c_0}$ is the asymptotic wave number for the pseudo-spin symmetry limit.

In order to solve Eq. (10) via the functional analytical method, we need to assume a wave function

$$G_{ps,n\kappa}(z) = z^\lambda (1-z)^{-\frac{ik_1}{2\alpha}} f(z), \tag{12}$$

with the pseudospin wave function parameter $\lambda = \frac{1}{2} + \frac{1}{2}\sqrt{1 + 4\kappa(\kappa-1) - \frac{4\beta D \sigma_0^2}{\alpha^2}}$. Inserting Eq. (12) into Eq. (10) leads to the formation of hypergeometric equation [53]

$$z(1-z)f''(z) + \left[2\lambda - \left(2\lambda - \frac{ik_1}{\alpha} + 1\right)z\right]f'(z) + \left[\left(\lambda - \frac{ik_1}{2\alpha}\right)^2 + P\right]f(z) = 0. \tag{13}$$

By considering the boundary condition that $f(z)$ tends to finite when $z \to 0$, the lower component of radial wave functions for any arbitrary $\kappa$−wave scattering states for the hyperbolical potential is obtained as [53]:

$$G_{ps,n\kappa}(z) = N_{n\kappa}(1 - e^{-2\alpha r})^\lambda e^{ik_1 r} {}_2F_1(a, b, c; 1 - e^{-2\alpha r}), \tag{14}$$

where

$$a = \lambda - \frac{ik_1}{2\alpha} - \sqrt{-\frac{\beta D \sigma_o}{\alpha^2} - \frac{k_1^2}{4\alpha^2}}, \quad b = \lambda - \frac{ik_1}{2\alpha} + \sqrt{-\frac{\beta D \sigma_o}{\alpha^2} - \frac{k_1^2}{4\alpha^2}}, \quad c = 2\lambda. \tag{15}$$

It is required that we consider the following conjugate relations which define the asymptotic phases

$$c - a - b = (a + b - c)^* = \frac{ik_1}{\alpha}, \tag{16a}$$

$$c - b = \lambda + \frac{ik_1}{2\alpha} - \sqrt{-\frac{\beta D \sigma_o}{\alpha^2} - \frac{k_1^2}{4\alpha^2}} = a^*, \tag{16b}$$

$$c - a = \lambda + \frac{ik_1}{2\alpha} + \sqrt{-\frac{\beta D \sigma_o}{\alpha^2} - \frac{k_1^2}{4\alpha^2}} = b^* \tag{16c}$$

and $N_{n\kappa}$ is the normalization.

### 3.1.2. Pseudospin symmetry phase shifts and normalization constant

To obtain the phase shifts $\delta_l$ and normalization constant, we apply the following recurrence relation of hypergeometric function or analytic-continuation formula [53]:

$${}_2F_1(a, b, c; z) = \frac{\Gamma(c)\Gamma(c-a-b)}{\Gamma(c-a)\Gamma(c-b)} {}_2F_1(a; b; 1 + a + b - c; 1 - z)$$

$$+(1-z)^{c-a-b}\frac{\Gamma(c)\Gamma(a+b-c)}{\Gamma a\Gamma(b)} {}_2F_1(c-a;\ c-b;\ c-a-b+1;\ 1-z). \tag{17}$$

Considering Eq. (17) and the property $_2F_1(a,b;\ c;0)=1$, as $r\to\infty$, we have

$$2F_1(a,b;\ c;1-e^{-2\alpha r}) = \Gamma(c)\left|\frac{\Gamma(c-a-b)}{\Gamma(c-a)\,\Gamma(c-b)} + \frac{\Gamma(c-a-b)}{\Gamma(a^*)\,\Gamma(b^*)}e^{-\alpha(c-a-b)r}\right|, \tag{18}$$

using Eq. (16), we may transform Eq. (18) as

$$2F_1(a,b;\ c;1-e^{-2\alpha r}) = \Gamma(c)\left|\frac{\Gamma(c-a-b)}{\Gamma(c-a)\,\Gamma(c-b)} + \left|\frac{\Gamma(c-a-b)}{\Gamma(c-a)\,\Gamma(c-b)}\right|^* e^{-ik_1 r}\right|. \tag{19}$$

By taking $\frac{\Gamma(c-a-b)}{\Gamma(c-a)\,\Gamma(c-b)} = \left|\frac{\Gamma(c-a-b)}{\Gamma(c-a)\,\Gamma(c-b)}\right|e^{i\delta}$ and inserting in Eq. (19), we have

$$2F_1(a,b;\ c;1-e^{-2\alpha r}) = \Gamma(c)\left|\frac{\Gamma(c-a-b)}{\Gamma(c-a)\,\Gamma(c-b)}\right|e^{-ik_1 r}\left[e^{i(k_1 r+\delta)} + e^{-i(k_1 r+\delta)}\right]. \tag{20}$$

Therefore, we obtain the asymptotic form of the lower spinor for $r\to\infty$ as

$$G_{ps,n\kappa}(r) = 2N_{n,\kappa}\Gamma(c)\left|\frac{\Gamma(c-a-b)}{\Gamma(c-a)\,\Gamma(c-b)}\right|\times \sin(k_1 r + \frac{\pi}{2} + \delta). \tag{21}$$

On comparison of Eq. (20) with the boundary condition $r\to\infty \Rightarrow G_{ps,n\kappa}(\infty)\to 2\sin\left(k_1 r + \delta_{l,n\kappa} - \frac{l\pi}{2}\right)$ [3]. Thus, we finally obtain the explicit pseudospin symmetry phase shifts and the normalization constant, respectively as;

$$\delta_{lps,n\kappa} = \frac{\pi}{2}(l+1) + arg\Gamma\left(\frac{ik_1}{\alpha}\right) - arg\Gamma\left(\lambda + \frac{ik_1}{2\alpha} + \sqrt{-\frac{\beta D\sigma_o}{\alpha^2} - \frac{k_1^2}{4\alpha^2}}\right) - arg\Gamma\left(\lambda + \frac{ik_1}{2\alpha} - \sqrt{-\frac{\beta D\sigma_o}{\alpha^2} - \frac{k_1^2}{4\alpha^2}}\right) \tag{22}$$

and

$$N_{ps,n\kappa} = \frac{\left|\Gamma\left(\lambda + \frac{ik_1}{2\alpha} + \sqrt{-\frac{\beta D\sigma_o}{\alpha^2} - \frac{k_1^2}{4\alpha^2}}\right)\right|}{\sqrt{2\lambda}} \times \left|\frac{\Gamma\left(\lambda + \frac{ik_1}{2\alpha} - \sqrt{-\frac{\beta D\sigma_o}{\alpha^2} - \frac{k_1^2}{4\alpha^2}}\right)}{\Gamma\left(\frac{ik_1}{\alpha}\right)}\right|. \tag{23}$$

### 3.1.3. Analytical properties of S-matrix for the pseudospin symmetry limit

Here, the analytical properties of partial-wave s-matrix is investigated to verify the fact that the poles of the s-matrix in the complex energy plane correspond to bound states for real poles [54], thus, we consider $\Gamma\left(\lambda + \frac{ik_1}{2\alpha} + \sqrt{-\frac{\beta D\sigma_o}{\alpha^2} - \frac{k_1^2}{4\alpha^2}}\right)$, where its first order poles is at the point [54]

$$\left(\lambda + \frac{ik_1}{2\alpha} + \sqrt{-\frac{\beta D\sigma_o}{\alpha^2} - \frac{k_1^2}{4\alpha^2}}\right) = 0, -1, -2, -3, \ldots = -n\ (n=0,1,2,\ldots). \tag{24}$$

Consequently, the bound state energy levels for the pseudospin symmetry limit is obtained as:

$$\frac{\gamma}{4\alpha^2} = \frac{\beta D}{4\alpha^2}(1-\sigma_0)^2 - \kappa(\kappa-1)c_0 + \left[\frac{(n+\lambda)^2 + \frac{\beta D\sigma_0}{\alpha^2}}{2(n+\lambda)}\right]^2. \tag{25}$$

## 3.2 Spin symmetry limit for the hyperbolical potential

Using the previously defined transformation variable and approximation, Eq. (8) becomes

$$\left\{\frac{d^2}{dz^2} - \frac{1}{(1-z)}\frac{d}{dz} + \left[\frac{\tilde{P}z^2+\tilde{Q}z+\tilde{R}}{z^2(1-z)^2}\right]\right\} F_{s,n\kappa}(r) = 0, \qquad (26)$$

with the following spin symmetry phase parameters:

$$\tilde{P} = \frac{k_2^2}{4\alpha^2} - \frac{\tilde{\beta}D\sigma_0}{\alpha^2}, \quad \tilde{Q} = \frac{4\alpha^2\kappa(\kappa+1)c_0 + 4\tilde{\beta}D\sigma_0(1+\sigma_0)}{4\alpha^2}, \quad \tilde{R} = \frac{-4\alpha^2\kappa(\kappa+1)c_0 - 4\tilde{\beta}D\sigma_0^2}{4\alpha^2}, \qquad (27)$$

where $k_2 = \sqrt{-\tilde{\beta}D(1-\sigma_0)^2 - \tilde{\gamma} - 4\alpha^2\kappa(\kappa+1)c_0}$ is the asymptotic wave number for the spin symmetry limit.

Similarly, we also assume the following upper wave function for the spin symmetry

$$F_{s,n\kappa}(z) = z^{\tilde{\lambda}}(1-z)^{-\frac{ik_2}{2\alpha}}f(z), \qquad (28)$$

with the spin symmetry wave function parameter $\tilde{\lambda} = \frac{1}{2} + \frac{1}{2}\sqrt{1 + 4\kappa(\kappa+1) - \frac{4\tilde{\beta}D\sigma_0^2}{\alpha^2}}$.

To avoid repetition, we follow the same procedures in previous sub-section and write the upper component of spin symmetry radial wave functions for any arbitrary $\kappa$ −wave scattering states as

$$F_{s,n\kappa}(z) = N_{n\kappa}(1 - e^{-2\alpha r})^{\tilde{\lambda}} e^{ik_2 r} {}_2F_1(a, b, c; 1 - e^{-2\alpha r}), \qquad (29)$$

where we have used the following wave function parameters

$$a = \tilde{\lambda} - \frac{ik_2}{2\alpha} - \sqrt{\frac{\tilde{\beta}D\sigma_0}{\alpha^2} - \frac{k_2^2}{4\alpha^2}}, \quad b = \tilde{\lambda} - \frac{ik_2}{2\alpha} + \sqrt{\frac{\tilde{\beta}D\sigma_0}{\alpha^2} - \frac{k_2^2}{4\alpha^2}}, \quad c = 2\tilde{\lambda}, \qquad (30)$$

where $N_{n\kappa}$ is the normalization constant depending on $n$ and $\kappa$.

### 3.2.2 Spin Symmetry Phase Shifts and Normalization Constant

Following the same steps in sub-Section **3.1.2**, we write the explicit spin symmetry phase shifts and the corresponding normalization constant, respectively as

$$\tilde{\delta}_{ls,n\kappa} = \frac{\pi}{2}(l+1) + arg\Gamma\left(\frac{ik_2}{\alpha}\right) - arg\Gamma\left(\tilde{\lambda} + \frac{ik_2}{2\alpha} + \sqrt{\frac{\tilde{\beta}D\sigma_0}{\alpha^2} - \frac{k_2^2}{4\alpha^2}}\right) - arg\Gamma\left(\tilde{\lambda} + \frac{ik_2}{2\alpha} - \sqrt{\frac{\tilde{\beta}D\sigma_0}{\alpha^2} - \frac{k_2^2}{4\alpha^2}}\right) \qquad (31)$$

and

$$N_{s,n\kappa} = \frac{\left|\Gamma\left(\tilde{\lambda} + \frac{ik_2}{2\alpha} + \sqrt{\frac{\tilde{\beta}D\sigma_0}{\alpha^2} - \frac{k_2^2}{4\alpha^2}}\right)\right|}{\sqrt{2\tilde{\lambda}}} \times \left|\frac{\Gamma\left(\tilde{\lambda} + \frac{ik_2}{2\alpha} - \sqrt{\frac{\tilde{\beta}D\sigma_0}{\alpha^2} - \frac{k_2^2}{4\alpha^2}}\right)}{\Gamma\left(\frac{ik_2}{\alpha}\right)}\right|, \qquad (32)$$

where we have employed the following phase shifts parameters for simplicity

$$c - a - b = (a + b - c)^* = \frac{ik_2}{\alpha}, \qquad (33)$$

$$c - b = \tilde{\lambda} + \frac{ik_2}{2\alpha} - \sqrt{\frac{\tilde{\beta}D\sigma_0}{\alpha^2} - \frac{k_2^2}{4\alpha^2}} = a^*, \quad (34)$$

$$c - a = \tilde{\lambda} + \frac{ik_2}{2\alpha} + \sqrt{\frac{\tilde{\beta}D\sigma_0}{\alpha^2} - \frac{k_2^2}{4\alpha^2}} = b^*. \quad (35)$$

### 3.2.3. Analytical Properties of S-matrix for the spin symmetry limit

Following the same fashion in sub-Section 3.13, the corresponding bound state energy levels for the spin symmetry are determined by the following energy equation:

$$\kappa(\kappa+1)c_0 + \frac{\tilde{\gamma}}{4\alpha^2} + \frac{\tilde{\beta}D}{4\alpha^2}(1-\sigma_0)^2 - \left[\frac{(n+\tilde{\lambda})^2 - \frac{\tilde{\beta}D\sigma_0}{\alpha^2}}{2(n+\tilde{\lambda})}\right]^2 = 0. \quad (36)$$

### 3.3. Non-relativistic limit for the scattering state solution

To study the non-relativistic limit, we apply the following appropriate mapping to Eq. (36):

$$\kappa(\kappa+1) = l(l+1), \quad E_{s,nk} - M \to E_{nl}, \quad M + E_{s,nk} \to \frac{2\mu}{\hbar^2}, \quad C_s = 0. \quad (37)$$

Consequently, we obtain the non-relativistic bound state energy levels for any arbitrary $l$ as:

$$E_{nl} = \frac{2\alpha^2\hbar^2 l(l+1)c_0}{\mu} + D(1-\sigma_0)^2 - \frac{\alpha^2\hbar^2}{8\mu}\left[\frac{(A+2n)^2 - \frac{8\mu D\sigma_0}{\alpha^2\hbar^2}}{(A+2n)}\right]^2, \quad (38)$$

where $\Lambda = 1 + \sqrt{1 + 4l(l+1) + \frac{8\mu D\sigma_0^2}{\alpha^2\hbar^2}}$.

**Table 1:** Pseudospin symmetry bound state energies at the poles of s-matrix (in units fm$^{-1}$) for the Hyperbolical potential as a function of positive potential parameter $\sigma_0$ for different value $n$ & $\kappa < 0$, $\alpha = 0.1$ in atomic units ($\mu = \hbar = 1$), $c_o = 0.0823058167837972$.

| $n$ | $\kappa$ | $\sigma_0$ | $E_{ps,n\kappa}$, $D = 5$, $C_{ps} = 0$ | $E_{ps,n,\kappa}$, $D = 10$, $C_{ps} = -5$ |
|---|---|---|---|---|
| 1 | -1 | 0.10 | 2.279123264, 1.029346029 | 2.123370278, -3.993494827 |
| 1 | -1 | 0.15 | 1.861708247, 1.031128744 | 1.321903847, -3.993385960 |
| 1 | -1 | 0.20 | 1.595741034, 1.034060836 | 0.841260815, -3.993229553 |
| 1 | -1 | 0.25 | 1.409954493, 1.038900136 | 0.513215975, -3.993021383 |
| 1 | -2 | 0.10 | 2.499877239, 1.049281052 | 2.396851228, -3.989095999 |
| 1 | -2 | 0.15 | 2.038561413, 1.051746477 | 1.492968827, -3.988960245 |
| 1 | -2 | 0.20 | 1.734612780, 1.055761011 | 0.958733145, -3.988765967 |
| 1 | -2 | 0.25 | 1.518422511, 1.062286830 | 0.598875815, -3.988508727 |
| 1 | -3 | 0.10 | 2.698880956, 1.074337203 | 2.742077063, -3.983652475 |
| 1 | -3 | 0.15 | 2.214535196, 1.077721454 | 1.721607908, -3.983480601 |
| 1 | -3 | 0.20 | 1.879494431, 1.083211792 | 1.120591256, -3.983235269 |
| 1 | -3 | 0.25 | 1.634306621, 1.092094144 | 0.719150716, -3.982911555 |
| 1 | -4 | 0.10 | 2.860137540, 1.104783298 | 3.119865055, -3.977163662 |
| 1 | -4 | 0.15 | 2.371800955, 1.109339737 | 1.987047112, -3.976948998 |
| 1 | -4 | 0.20 | 2.015203200, 1.116738148 | 1.314844872, -3.976643120 |
| 1 | -4 | 0.25 | 1.745286576, 1.128746703 | 0.866633979, -3.976240458 |
| 2 | -1 | 0.10 | 2.615024144, 1.055156519 | 3.119164186, -3.987967157 |
| 2 | -1 | 0.15 | 2.170905660, 1.059554992 | 2.174347963, -3.987685287 |
| 2 | -1 | 0.20 | 1.854762182, 1.066940767 | 1.576374701, -3.987279697 |

| n | κ | $\sigma_0$ | $E_{n, \kappa<0}, D=10, C_s=0$ | $E_{n, \kappa<0}, D=10, C_s=5$ |
|---|---|---|---|---|
| 2 | -1 | 0.25 | 1.610620539, 1.079632694 | 1.152943758, -3.986738757 |
| 2 | -2 | 0.10 | 2.751526391, 1.082577001 | 3.308028013, -3.982028646 |
| 2 | -2 | 0.15 | 2.286314037, 1.087870608 | 2.296760801, -3.981727996 |
| 2 | -2 | 0.20 | 1.948130591, 1.096737249 | 1.662607037, -3.981295986 |
| 2 | -2 | 0.25 | 1.684753900, 1.111945302 | 1.217121129, -3.980720876 |
| 2 | -3 | 0.10 | 2.880207106, 1.115665997 | 3.556476381, -3.975030556 |
| 2 | -3 | 0.15 | 2.407058769, 1.122328052 | 2.465317807, -3.974688096 |
| 2 | -3 | 0.20 | 2.050146916, 1.133485527 | 1.784093810, -3.974197233 |
| 2 | -3 | 0.25 | 1.766570973, 1.152692476 | 1.308784059, -3.973545930 |
| 2 | -4 | 0.10 | 2.984654813, 1.154764895 | 3.839051072, -3.966978840 |
| 2 | -4 | 0.15 | 2.516213585, 1.163231610 | 2.667142308, -3.966583110 |
| 2 | -4 | 0.20 | 2.146431337, 1.177478350 | 1.933530119, -3.966017016 |
| 2 | -4 | 0.25 | 1.843996674, 1.202334754 | 1.423425408, -3.965267870 |

**Table 2:** Spin symmetry energies at the poles of s-matrix (in units fm$^{-1}$) for the Hyperbolical potential as a function of positive potential parameter $\sigma_0$ for different value $n$ and $\kappa < 0$, $\alpha = 0.1$ & $D = 10$ in atomic units ($\mu = \hbar = 1$), $c_o = 0.0823058167837972$.

| n | κ | $\sigma_0$ | $E_{n, \kappa<0}, D=10, C_s=0$ | $E_{n, \kappa<0}, D=10, C_s=5$ |
|---|---|---|---|---|
| 0 | -2 | 0.10 | 2.852748850, -0.997190645 | 4.894455875, 4.004884016 |
| 0 | -2 | 0.15 | 2.286237200, -0.997160056 | 4.484630891, 4.005026090 |
| 0 | -2 | 0.20 | 1.976219709, -0.997116299 | 4.291441866, 4.005242281 |
| 0 | -2 | 0.25 | 1.777900928, -0.997058390 | 4.186232796, 4.005555499 |
| 0 | -3 | 0.10 | 3.349197922, -0.994101445 | 5.374102366, 4.010243686 |
| 0 | -3 | 0.15 | 2.597824107, -0.994048465 | 4.789547550, 4.010499561 |
| 0 | -3 | 0.20 | 2.189987655, -0.993972934 | 4.496203764, 4.010885663 |
| 0 | -3 | 0.25 | 1.933180571, -0.993873429 | 4.328714233, 4.011438071 |
| 0 | -4 | 0.10 | 3.898927388, -0.989885456 | 5.855300432, 4.017565178 |
| 0 | -4 | 0.15 | 2.970569858, -0.989803165 | 5.119359731, 4.017971982 |
| 0 | -4 | 0.20 | 2.458487266, -0.989686044 | 4.729675556, 4.018583410 |
| 0 | -4 | 0.25 | 2.134995974, -0.989532099 | 4.497834585, 4.019453088 |
| 0 | -5 | 0.10 | 4.444897842, -0.984540394 | 6.309088175, 4.026860903 |
| 0 | -5 | 0.15 | 3.364895150, -0.984421967 | 5.450837766, 4.027456271 |
| 0 | -5 | 0.20 | 2.754447857, -0.984253582 | 4.974848046, 4.028349215 |
| 0 | -5 | 0.25 | 2.364312495, -0.984032549 | 4.681673312, 4.029615355 |
| 1 | -2 | 0.10 | 4.307215786, -0.992940748 | 5.855244196, 4.012347926 |
| 1 | -2 | 0.15 | 3.494784849, -0.992807827 | 5.188936272, 4.012901328 |
| 1 | -2 | 0.20 | 3.001592356, -0.992615929 | 4.804784938, 4.013761503 |
| 1 | -2 | 0.25 | 2.661226629, -0.992358819 | 4.556938740, 4.015047441 |
| 1 | -3 | 0.10 | 4.595818435, -0.988170213 | 6.178579520, 4.020644486 |
| 1 | -3 | 0.15 | 3.680400729, -0.988003000 | 5.414210564, 4.021381830 |
| 1 | -3 | 0.20 | 3.132061101, -0.987762645 | 4.970596099, 4.022516492 |
| 1 | -3 | 0.25 | 2.758336616, -0.987442474 | 4.682895638, 4.024188218 |
| 1 | -4 | 0.10 | 4.953368438, -0.982265799 | 6.534179272, 4.030932630 |
| 1 | -4 | 0.15 | 3.924033854, -0.982052870 | 5.675017706, 4.031906704 |
| 1 | -4 | 0.20 | 3.308761019, -0.981747682 | 5.167809668, 4.033396291 |
| 1 | -4 | 0.25 | 2.892498344, -0.981342717 | 4.835280862, 4.035570778 |
| 1 | -5 | 0.10 | 5.338258842, -0.975226530 | 6.885351175, 4.043223541 |
| 1 | -5 | 0.15 | 4.201954406, -0.974959459 | 5.945777082, 4.044479104 |
| 1 | -5 | 0.20 | 3.517066442, -0.974577405 | 5.378157524, 4.046391514 |
| 1 | -5 | 0.25 | 3.054118376, -0.974071753 | 5.000762561, 4.049166887 |

**Table 3:** Non-relativistic energies at the poles of s-matrix (in units fm$^{-1}$) for the Hyperbolical potential as a function of positive potential parameter $\sigma_0$ for different states in atomic units ($\mu = \hbar = 1$), $c_o = 0.0823058167837972$ and $D = 10$ in all the calculations

| n | l | $\sigma_0$ | States | $E_{nl}$ for $\alpha = 0.10$ | $E_{nl}$ for $\alpha = 0.15$ | $E_{nl}$ for $\alpha = 0.20$ | $E_{nl}$ for $\alpha = 0.25$ |
|---|---|---|---|---|---|---|---|
| 0 | 1 | 0.10 | 2p | 2.61886 | 3.90571 | 5.00379 | 5.88669 |
|   |   | 0.15 |    | 1.68039 | 2.57787 | 3.43316 | 4.20997 |
|   |   | 0.20 |    | 1.20888 | 1.86663 | 2.52048 | 3.14740 |
| 1 | 1 | 0.10 | 3p | 4.73552 | 6.04570 | 6.91711 | 7.48475 |
|   |   | 0.15 |    | 3.46026 | 4.62307 | 5.50067 | 6.15044 |
|   |   | 0.20 |    | 2.68320 | 3.67154 | 4.46564 | 5.09305 |
| 0 | 2 | 0.10 | 3d | 3.62734 | 5.29485 | 6.47635 | 7.25747 |
|   |   | 0.15 |    | 2.27011 | 3.56704 | 4.69665 | 5.59830 |
|   |   | 0.20 |    | 1.57908 | 2.54853 | 3.48262 | 4.31329 |
| 2 | 1 | 0.10 | 4p | 6.00299 | 7.11553 | 7.71951 | 8.02106 |
|   |   | 0.15 |    | 4.66770 | 5.80657 | 6.52463 | 6.95731 |
|   |   | 0.20 |    | 3.75704 | 4.81242 | 5.53159 | 6.00361 |
| 1 | 2 | 0.10 | 4d | 5.33164 | 6.73663 | 7.54623 | 7.97844 |
|   |   | 0.15 |    | 3.85813 | 5.19480 | 6.13553 | 6.75588 |
|   |   | 0.20 |    | 2.95293 | 4.10491 | 5.00322 | 5.66533 |
| 0 | 3 | 0.10 | 4f | 4.69036 | 6.43153 | 7.43683 | 7.97990 |
|   |   | 0.15 |    | 3.00341 | 4.60144 | 5.79817 | 6.60873 |
|   |   | 0.20 |    | 2.07413 | 3.35783 | 4.47694 | 5.35270 |

**Table 4:** Pseudospin scattering phase shifts for Hyperbolical potential with positive potential parameter $\sigma_0$=0.10, $\alpha = 0.10$, $c_o = 0.0823058167837972$ and $E = 1$. The relativistic mass is m=1 in atomic units ($m = \hbar = 1$).

| l | κ | $\delta_{l,ps,\kappa}$ for $C_{ps} = 0, D = 10$ | $\delta_{l,ps,\kappa}$ for $C_{ps} = 0.05, D = 10$ |
|---|---|---|---|
| 0 | -1, 1 | 1.523874403924851, 1.570796326794897 | -8.078248081182888, -6.037944299866905 |
|   | -2, 2 | 0.458609436071745, 1.523874403924851 | -9.297898988297444, -8.078248081182888 |
|   | -3, 3 | -1.072342966122305, 0.458609436071745 | -9.919223591558044, -9.297898988297444 |
|   | -4, 4 | -2.932321514421307, -1.072342966122305 | -10.040884779950039, -9.919223591558044 |
|   | -5, 5 | -5.047233964874891, -2.932321514421307 | -9.667030052888922, -10.040884779950039 |
| 1 | -1, 1 | 3.094670730719748, 3.141592653589793 | -6.507451754387993, -4.467147973072009 |
|   | -2, 2 | 2.029405762866642, 3.094670730719748 | -7.727102661502547, -6.507451754387993 |
|   | -3, 3 | 0.498453360672592, 2.029405762866642 | -8.348427264763147, -7.727102661502547 |
|   | -4, 4 | -1.361525187626411, 0.498453360672592 | -8.470088453155141, -8.348427264763147 |
|   | -5, 5 | -3.476437638079995, -1.361525187626411 | -8.096233726094024, -8.470088453155141 |
| 2 | -1, 1 | 4.665467057514643, 4.712388980384690 | -4.936655427593096, -2.896351646277112 |
|   | -2, 2 | 3.600202089661539, 4.665467057514643 | -6.156306334707651, -4.936655427593096 |
|   | -3, 3 | 2.069249687467488, 3.600202089661539 | -6.777630937968251, -6.156306334707651 |
|   | -4, 4 | 0.209271139168486, 2.069249687467488 | -6.899292126360244, -6.777630937968251 |
|   | -5, 5 | -1.905641311285098, 0.209271139168486 | -6.525437399299129, -6.899292126360244 |
| 3 | -1, 1 | 6.236263384309540, 6.283185307179586 | -3.365859100798200, -1.325555319482215 |
|   | -2, 2 | 5.170998416456435, 6.236263384309540 | -4.585510007912754, -3.365859100798200 |
|   | -3, 3 | 3.640046014262385, 5.170998416456435 | -5.206834611173354, -4.585510007912754 |
|   | -4, 4 | 1.780067465963383, 3.640046014262385 | -5.328495799565348, -5.206834611173354 |
|   | -5, 5 | -0.334844984490202, 1.780067465963383 | -4.954641072504233, -5.328495799565348 |

Table 5: Pseudospin scattering phase shifts for Hyperbolical potential with positive potential parameter $\sigma_0$=0.50, $\alpha = 0.50$, $c_o = 0.0823058167837972$ and $E = 1$. The relativistic mass is m=1 in atomic units ($m = \hbar = 1$).

| $l$ | κ | $\delta_{l,s,\kappa}$ for $C_s = 5$, $D = 10$ | $\delta_{l,s,\kappa}$ for $C_s = 10$, $D = 10$ |
|---|---|---|---|
| 0 | -1, 1 | -15.357449458632775, -15.180768491483812 | -34.356717558868027, -34.213439256222195 |
|   | -2, 2 | -15.180768491483812, -14.812590524986316 | -34.213439256222195, -33.922995271131597 |
|   | -3, 3 | -14.812590524986316, -14.223834178887669 | -33.922995271131597, -33.477658101510812 |
|   | -4, 4 | -14.223834178887669, -13.370909470812821 | -33.477658101510812, -32.865901014031351 |
|   | -5, 5 | -13.370909470812821, -12.192227544491045 | -32.865901014031351, -32.072333895105686 |
| 1 | -1, 1 | -13.786653131837877, -13.609972164688918 | -32.785921232073129, -32.642642929427303 |
|   | -2, 2 | -13.609972164688918, -13.241794198191421 | -32.642642929427303, -32.352198944336699 |
|   | -3, 3 | -13.241794198191421, -12.653037852092771 | -32.352198944336699, -31.906861774715914 |
|   | -4, 4 | -12.653037852092771, -11.800113144017923 | -31.906861774715914, -31.295104687236456 |
|   | -5, 5 | -11.800113144017923, -10.621431217696150 | -31.295104687236456, -30.501537568310784 |
| 2 | -1, 1 | -10.645060478248087, -10.468379511099124 | -31.215124905278238, -31.071846602632405 |
|   | -2, 2 | -10.468379511099124, -10.100201544601624 | -31.071846602632405, -30.781402617541801 |
|   | -3, 3 | -10.100201544601624, -9.5114451985029810 | -30.781402617541801, -30.336065447921015 |
|   | -4, 4 | -9.5114451985029810, -8.6585204904281290 | -30.336065447921015, -29.724308360441558 |
|   | -5, 5 | -8.6585204904281290, -7.4798385641063540 | -29.724308360441558, -28.930741241515886 |
| 3 | -1, 1 | -10.645060478248087 -10.468379511099124 | -29.644328578483339, -29.501050275837507 |
|   | -2, 2 | -10.468379511099124, -10.100201544601624 | -29.501050275837507, -29.210606290746910 |
|   | -3, 3 | -10.100201544601624, -9.5114451985029810 | -29.210606290746910, -28.765269121126117 |
|   | -4, 4 | -9.5114451985029810, -8.6585204904281290 | -28.765269121126117, -28.153512033646667 |
|   | -5, 5 | -8.6585204904281290, -7.4798385641063540 | -28.153512033646667, -27.359944914720987 |

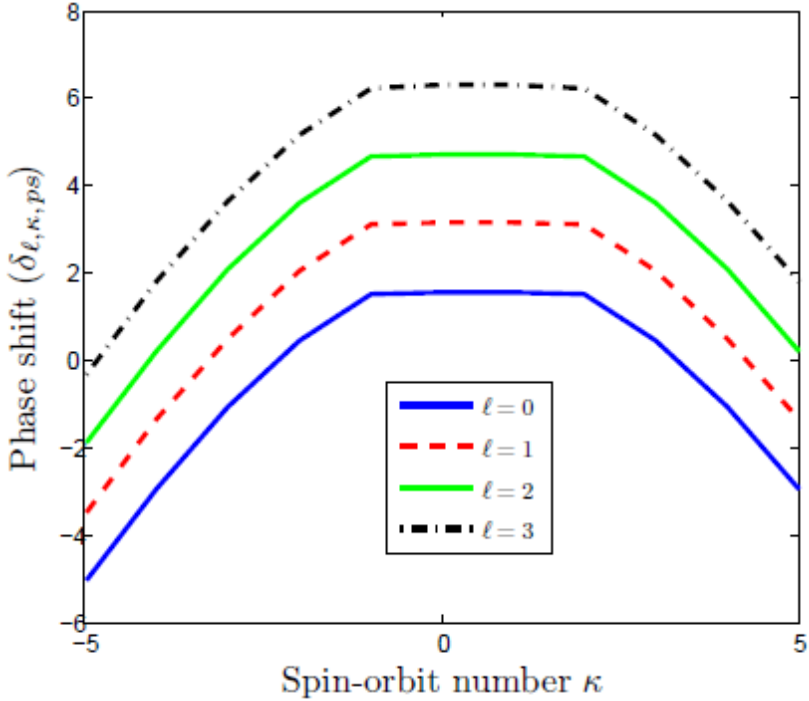

Figure 1: A plot of pseudospin scattering phase shifts for the Hyperbolical potential as a function of spin-orbit number κ for $l = 0, 1, 2, 3$ with positive potential parameter $\sigma_0$=0.10, $\alpha = 0.10$, $c_o = 0.0823058167837972$, $C_{ps} = 0$, $D = 10$ and $E = M = 1$. The relativistic mass and energy are equal in all the calculations.

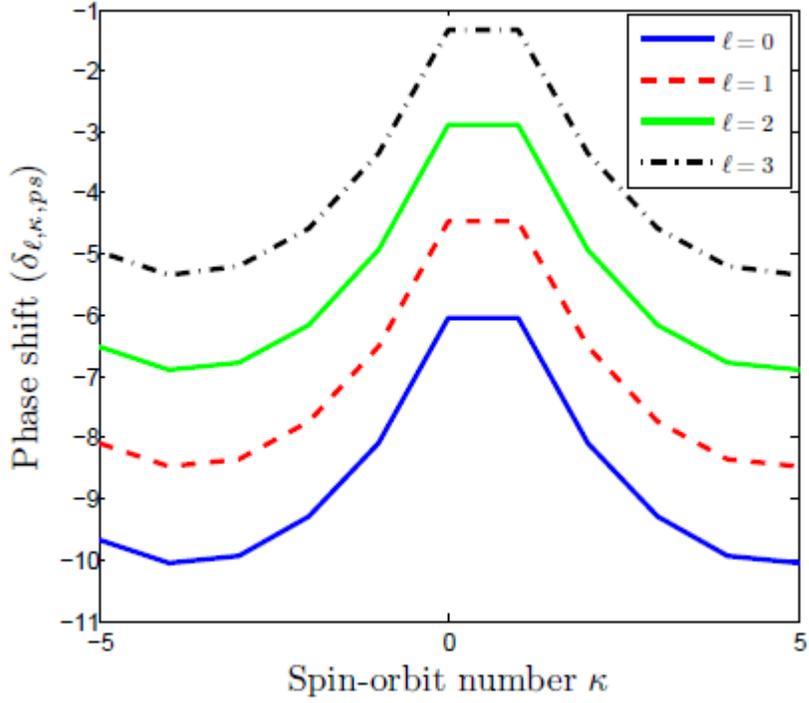

Figure 2: A plot of pseudospin scattering phase shifts for the Hyperbolical potential as a function of spin-orbit number κ for $l = 0, 1, 2, 3$ with positive potential parameter $\sigma_0$=0.10, $\alpha = 0.10$, $c_o = 0.0823058167837972$, $C_{ps} = 0.05$, $D = 10$ and $E = M = 1$. The relativistic mass and energy are equal in all the calculations.

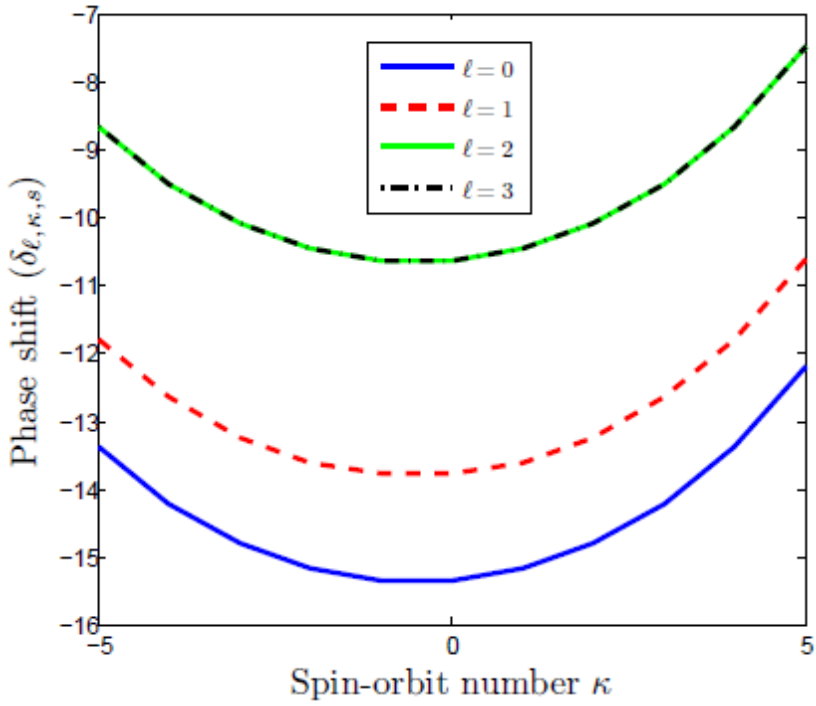

Figure 3: A plot of spin scattering phase shifts for the Hyperbolical potential as a function of spin-orbit number κ for $l = 0, 1, 2, 3$ with positive potential parameter $\sigma_0$=0.50, $\alpha = 0.50$, $c_o = 0.0823058167837972$, $C_s = 5$, $D = 10$ and $E = M = 1$. The relativistic mass and energy are equal in all the calculations.

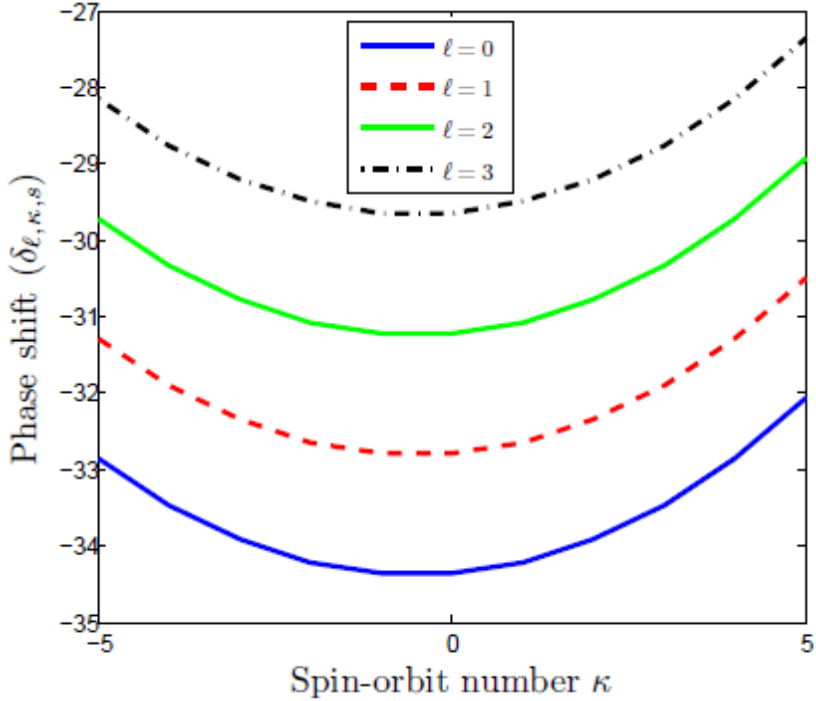

Figure 4: A plot of spin scattering phase shifts for the Hyperbolical potential as a function of spin-orbit number κ for $l = 0, 1, 2, 3$ with positive potential parameter $\sigma_0$=0.50, $\alpha = 0.50$, $c_o = 0.0823058167837972$, $C_s = 10$, $D = 10$ and $E = M = 1$. The relativistic mass and energy are equal in all the calculations.

## 5. Discussion and Conclusion

The pseudospin symmetry bound states energy spectra display in table 1 is obtained from Eq. (25) and the corresponding spin symmetry bound states energy spectra display in table 2 is obtained from Eq. (36) while the non-relativistic bound state energy spectra is obtained from Eq. (38). The pseudospin symmetry and spin symmetry phase shifts are obtained from Eq. (22) and Eq. (31) respectively.

In Tables 1 and 2, for a fixed values principal quantum numbering values $n$, the relativistic bound state energies increase with decreasing of spin –orbit quantum numbers κ whether the symmetry constants are present or not. The results reasonably showed that the presence symmetry constants contribute significantly to the relativistic bound state energies. Table 3 displayed non-relativistic energies for the hyperbolical potential which is in excellent agreement with the one obtained by asymptotic iteration method [45] for the state 2p, 3p, 3d, 4p, 4d and 4f for positive potential parameter $\sigma_0 = 0.10$.

The pseudospin symmetry and spin symmetry phase shifts are displayed in table 4 and table 5 respectively. To see the clearer behaviour of the phase shifts in the spin and pseudospin symmetry limits we plot the numerical results of phase shift in both pseudospin and spin symmetries against the spin-orbit quantum numbers κ in the figures 1-4. In figure 1, Phase shifts is slightly decreasing exponentially to the left and to the right for negative and positive values of spin-orbit quantum numbers, respectively for zero pseudo spin constant. A steady phase shifts is observed at $-1 \leq \kappa \leq 1$. This suggest that the spin-orbit number affect phase shifts significantly for any arbitrary angular momentum quantum number.

Figure 2 illustrate the behaviour of phase shifts in the presence of pseudo spin constants, the graphs follow the same pattern with a more negative phase shifts. The negativity is an indication that pseudospin constant strongly influence scattering phase shifts. However, Figure 3 and 4 illustrate the behaviour of spin symmetry phase shifts. An exponential rise in phase shifts to the left and to the right for negative and positive value of the spin-orbit quantum numbers κ, respectively is observed for all $l$. In figure 3, an overlapping of phase shifts is observed for $l = 2$ &3 indicating that the particle of these angular momentum number will experience the same phase shifts. A reasonable increase in spin symmetry constant give a more distinctive phase shifts (see figure 4).

In conclusion, we have studied the approximate scattering state solution of Dirac equation with the hyperbolical potential using a short-range approximation within the framework of functional analytical method. We have obtained the spin and pseudosin symmetry bound state energies and their corresponding non-relativistic energies, spin and pseudospin symmetry phase shifts, normalization constants, pseudospin symmetry lower component and spin symmetry upper component of radial spinor wave functions for any arbitrary $\kappa$ −wave scattering states.

We also studied the behaviour of phase shifts with spin-orbit quantum numbers under spin and pseudo spin symmetries and we have successfully showed that relativistic scattering phase shifts is largely depend on the symmetry constants ($C_{ps}$ and $C_s$) and spin-orbit quantum numbers $\kappa$

**Acknowledgements**

Oluwadare O.J declared that this paper is an output of my Ph.D. research thesis supported by the Tertiary Education Trust Funds (TETFunds) through the Federal University Oye Ekiti, Ekiti State, Nigeria.